\documentclass[pra, aps, twocolumn, groupedaddress,  superscriptaddress]{revtex4}
\usepackage{slashed}
\usepackage{graphicx}
\usepackage{subfigure}
\usepackage[usenames, dvipsnames]{color}
\usepackage{graphics}
\usepackage{hyperref}
\usepackage{bm}
\usepackage{amsfonts}

\usepackage{amsmath}

\hypersetup{backref,
colorlinks=true,
urlcolor=blue,
linkcolor=blue,
linktoc=page,
citecolor=blue}

\everymath{\displaystyle} 

\begin{document}

\title{ 
High-Fidelity Quantum Information Transmission Using a Room-Temperature Nonrefrigerated Lossy Microwave Waveguide
}


\author{Montasir Qasymeh$^{*}$}
\affiliation{Electrical and Computer Engineering Department, Abu Dhabi University, 59911 Abu Dhabi, UAE}
\author{Hichem Eleuch}
\affiliation{Department of Applied Physics and Astronomy, University of Sharjah, Sharjah, United Arab Emirates}
\affiliation{Institute for Quantum Science and Engineering, Texas AM University, College Station, TX 77843 USA}

\begin{abstract}
Quantum microwave transmission is key to realizing modular superconducting quantum computers \cite{logic1}  and distributed quantum networks\cite{Deterministic2}. A large number of incoherent photons are thermally generated within the microwave frequency spectrum. The closeness of the transmitted quantum state to the source-generated quantum state at the input of the transmission link (measured by the transmission fidelity) degrades due to the presence of the incoherent photons. Hence, high-fidelity quantum microwave transmission has long been considered to be infeasible without refrigeration\cite{Microwave3,coherent4}. In this study, we propose a novel method for high-fidelity quantum microwave transmission using a room-temperature lossy waveguide. The proposed scheme consists of connecting two cryogenic nodes (i.e., a transmitter and a receiver) by the room-temperature lossy microwave waveguide. First, cryogenic preamplification is implemented prior to transmission. Second, at the receiver side, a cryogenic loop antenna is placed inside the output port of the waveguide and coupled to an \textit{LC} harmonic oscillator located outside the waveguide. The loop antenna converts quantum microwave fields to a quantum voltage across the coupled \textit{LC} harmonic oscillator. Noise photons are induced across the \textit{LC} oscillator including the source generated noise, the preamplification noise, the thermal occupation of the waveguide, and the fluctuation-dissipation noise. The loop antenna detector at the receiver is designed to extensively suppress the induced photons across the \textit{LC} oscillator. The signal transmittance is maintained intact by providing significant preamplification gain. Our calculations show that high-fidelity quantum transmission (i.e., more than $95\%$)  is realized based on the proposed scheme for transmission distances reaching 100 m. 
\end{abstract}
\maketitle

\section{Introduction}
Realizing large-scale quantum computers with thousands (or millions) of qubits requires efficient quantum data transmission between distant quantum nodes \cite{Demonstrating5,information6,Distributed7}. This architecture of remotely connected quantum modules is known as a modular quantum computer \cite{modular8}, which has been purported as a means of overcoming the current challenges that prevent the scale-up of quantum computers, such as crosstalk, input/output coupling limitations, and limited space \cite{Engineering9,computers10}. Likewise, future quantum sensing applications and networks require efficient quantum transmissions with applicable implementations \cite{Interconnects11,internet12,internet13,Entanglement14,Sensing15}. Among the main quantum technologies that have been developed, superconducting-based quantum circuits have shown exceptional potential for quantum signal processing and computation \cite{Josephson16,supremacy17,ours18}. However, superconducting signals operate in the microwave frequency spectrum and are therefore particularly vulnerable to degradation by thermal energy. This is one of the reasons for housing superconducting circuits in cryostats. Such operating condition imposes strict limitations on the ability to build modular superconducting quantum computers. Several approaches have been proposed to connect distant superconducting quantum circuits. One approach is based on entangling distant superconducting circuits using coaxial cables carrying microwave photons \cite{memories19,Deterministic20,Deterministic21} or acoustic channels carrying phonons \cite{quantum22}. Transmission lengths between 1 and 2 m have been reported using this technique. Another approach involves cooling a microwave waveguide to cryogenic temperatures \cite{Microwave3}. Five meter coherent microwave transmission was reported. These two approaches require housing transmission channels to be placed in dilution refrigerators, which is challenging in terms of both economy and implementation. In this regard, IBM has announced plans to build a gigantic liquid-helium refrigerator, 10 feet tall and 6 feet wide, to support a 1000 qubit quantum computer that is planned for construction in 2023 and a milestone million qubit quantum computer that is planned for construction in 2030 \cite{IBM23}. 
Quantum information transmission via noisy channels was proposed using a time-dependent coupling between qubits (at the transmitter and receiver) and a connecting channel \cite{thermalch1,thermalch2}. However, channel loss and dissipation-generated noise were not taken into account. This approach requires operating at a few Kelvins (i.e., $4$ K) and implementing quantum-error correction to attain high-fidelity transmission. Other researchers have proposed alternative approaches in which optical fibers are used to connect superconducting cryogenic circuits (or processors), facilitated by microwave-to-optical transduction \cite{coherent4,our24,patent25,Superconducting26}. However, several challenges remain in realizing efficient wide-band electro-optic transducers, and drawbacks are associated with transduction resulting from the addition of conversion noise and laser-induced quasiparticles\cite{Microwave3}. 

In this study, we propose a novel approach for transmitting coherent quantum microwave fields using a room-temperature lossy microwave waveguide. In the proposed scheme, two distant superconducting circuits housed in cryostats are connected by a microwave waveguide that is placed outside the refrigerators and operates at room temperature, as shown in Fig. 1. 
\begin{figure}
\centering
\includegraphics[width=1\linewidth]{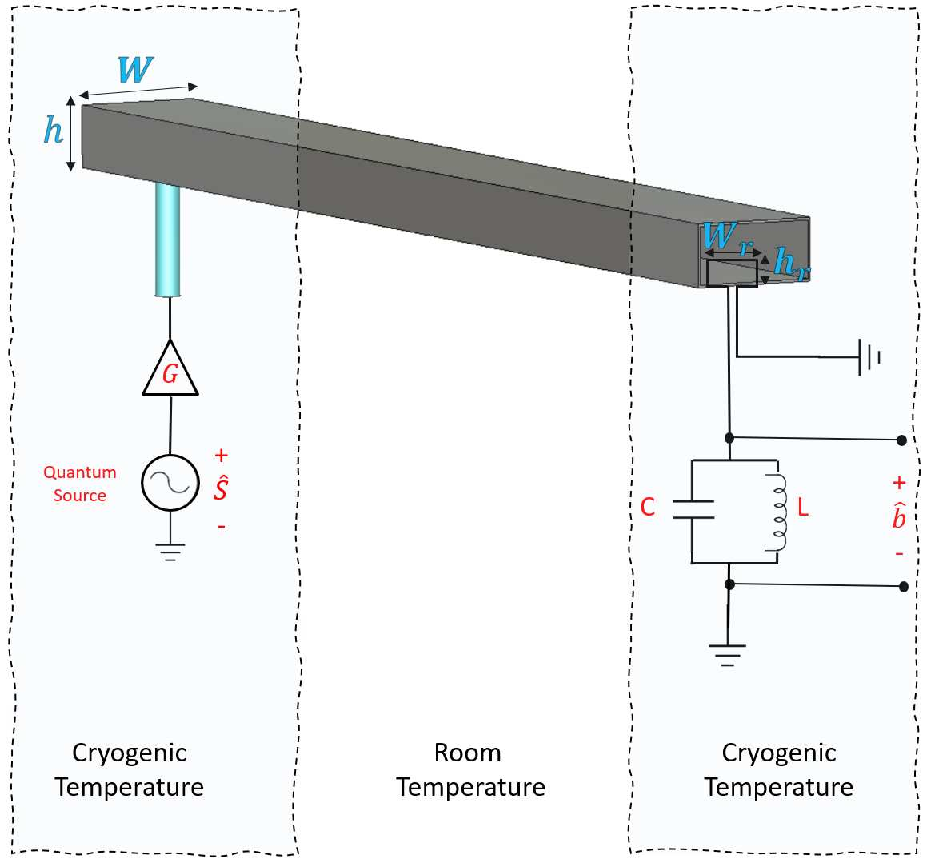}
\caption{Proposed system for microwave transmission using a room-temperature waveguide.}
\label{fig:frog}
\end{figure}
Our approach implies implementing a mechanism for photons suppression at the input port of the cryogenic receiver. It then follows that the noise photons are reduced substantially by imposing intensive suppression. However, two conditions are required to conduct high-fidelity information transmission. The first condition is to attain the suppression process while inducing no additional noise photons (e.g., as opposed to attenuation). The second condition is to preserve a significant number of signal photons despite suppression. To this end, we propose utilizing a superconducting loop antenna coupled to an \textit{LC} harmonic oscillator at the receiving end of the transmission waveguide. According to \textit{Faraday's law} of induction, the loop antenna converts microwave fields to microwave voltages across the \textit{LC} harmonic oscillator. Voltages are induced in the \textit{LC} oscillator associated with both the transmitted signal and the noise photons. Interestingly, the use of a loop antenna with suitably designed dimensions can significantly suppress the number of induced noise photons in the \textit{LC} harmonic oscillator. Moreover, cryogenic preamplification at the transmitter side can be used to maintain the signal's transmittance close to unity. Importantly, the added preamplification noise are subjected to the waveguide attenuation and the loop antenna suppression. Consequently, we show that near-quantum-limited noise temperature transmission can be achieved. The principle of operation is based on intensifying (amplifying) the quantum fields before transmission. It then follows that the propagating fields are immune to the losses and the added noise. The suppression (at the output port of the transmission channel) provides two important functionalities. First, the received fields are scaled back to the quantum level. Second, the accumulating noise photons are suppressed. The crucial key is to implement the amplification and the suppression processes without degrading the purity of the transmitted quantum signal. The concept of transmission-fidelity can be employed to quantify the impact of added incoherent photons during the amplification, transmission, and suppression processes. For instance, the closeness of the quantum state of the received qubit to the quantum state of the transmitted qubit is measured by the transmission fidelity. Our calculations show a transmission-fidelity of about 95 \%  can be achieved by using our proposed system for up to 100 m transmission distances. Consequently, the fundamental quantum processes are naturally inherited by the transmitted quantum state, with the advantage of extending the interaction over distances. For instance, similar to the case of the cryogenically refrigerated microwave waveguide in ref. \cite{Microwave3}, the transmitted quantum state can establish a remote entanglement of distant qubits. Hence, the proposed scheme enables the use of a microwave waveguide operating at room temperature to connect two superconducting quantum circuits housed in distantly separated cryostats. This approach has the potential to realize a modular superconducting quantum computer (or a local quantum network) without transduction or waveguide cooling. While the purpose of the current work is to provide a viable scheme for coherent transmission,  the functionality of the proposed system is another aspect that is beyond the scope of this work.

The organization of the paper is as in the following. In section II, the model of the proposed transmission system considering a lossy room-temperature waveguide is developed. The associated fields and the equation of motion are derived in subsection (A). Subsection (B) presents the quantum signal and noise operators. The system's performance is evaluated in section III. The effective noise temperature and the qubit transmission-fidelity are modeled and calculated considering a typical aluminum waveguide. Finally, section IV discusses the future potential of the proposed system and highlights its economic and practical advantages.
\section{System}
Consider a system of two separated superconducting quantum nodes (transmitter and a receiver) placed in two distant dilution refrigerators. The two nodes are connected by a nonrefrigerated lossy microwave waveguide, as shown in Fig. 1. Here, $W$ and $h$ are the width and the height of the waveguide along the $x$ and $y$ axes, respectively. 
\subsection{Propagating Fields}
A typical rectangular microwave waveguide is implemented for signal transmission. The waveguide is  supporting the fundamental $TE_{10}$ mode. Hence, the associated electric and magnetic fields can be expressed as follows \cite{Pozar32}:
\begin{equation} \label{m1}
\vec{E}(x,y,x,t)=i A Z_F \Omega sin(\frac{\pi x}{W})e^{(i\beta z-\omega t)} \vec{e}_y+c.c.,
\end{equation}

\begin{equation} \label{m2}
\begin{split}
\vec{H}(x,y,z,t)& =-i A \sqrt{\Omega^2-1} sin(\frac{\pi x}{W})e^{(i\beta z-\omega t)} \vec{e}_x \\
 &+ A \sqrt{\Omega^2-1} sin(\frac{\pi x}{W})e^{(i\beta z-\omega t)} \vec{e}_z+c.c.,
\end{split}
\end{equation}
where $A$ is the complex amplitude of the $TE_{10}$ mode,  $Z_F=377\sqrt{\frac{\mu_r}{\epsilon_r}}$ represents the impedance of the filling material, and $\Omega=\frac{\omega}{\omega_c}$. Here, $\omega$ is the microwave signal frequency; $\omega_c=\frac{2\pi c}{2W\sqrt{\epsilon_r}}$ denotes the cut-off frequency; $\mu_r$ and $\epsilon_r$ are the relative permeability and permittivity of the filling material, respectively; $\beta$ represents a propagation constant; and $c$ is the speed of light in vacuum.

The classical Hamiltonian of the $TE_{10}$ mode is given by $\mathcal{H}=\frac{1}{2}\epsilon_0 \epsilon_{eff} \mid A \mid^2 Z_F^2 \Omega^2 V_{ol}+\frac{1}{2}\mu_0 \mu_{r} \mid A \mid^2 \Omega^2 V_{ol}$, where $V_{ol}=W\times h\times l$ denotes the waveguide volume and $l$ is the waveguide length. The propagating microwave field can be quantized through the following relation:
\begin{equation} \label{m3}
   A=\frac{(\hbar \omega)^\frac{1}{2}}{\varphi^\frac{1}{2}(\epsilon_0\epsilon_{eff} V_{ol})^\frac{1}{2}} \hat{a},
\end{equation}
where $\hat{a}$ represents the annihilation operator of the $TE_{10}$ mode, $\varphi=\frac{\Omega^2 Z_F^2}{2}+\frac{\mu_0 \mu_r \Omega^2}{2\epsilon_0 \epsilon_{eff}}$, and $\epsilon_{eff}=\epsilon_r-\frac{\pi^2 c^2}{W^2 \omega^2}$ denotes the effective permittivity of the waveguide. Thus, the quantum Hamiltonian is given by $\hat{\mathcal{H}}=\hbar \omega \hat{a}^\dagger \hat{a}$.

The equation of motion can be obtained by substituting the quantum Hamiltonian into the Heisenberg equation $\frac{\partial \hat{a} }{\partial t}=\frac{i}{\hbar}[\hat{\mathcal{H}},\hat{a}]$. Incorporating the waveguide dissipation and the dissipation-fluctuation noise into this equation, yields $\frac{\partial\hat{a}}{\partial t}=-i\omega\hat{a}-\frac{\Gamma}{2} \hat{a}+\sqrt{\Gamma}\hat{f}_L$. Consequently, by using the rotation approximation and setting $\hat{a}=\hat{u} e^{-i\omega t}$, the equation of motion is given in the following form: \begin{equation} \label{m5}
\frac{\partial\hat{u}}{\partial t}=-\frac{\Gamma}{2} \hat{u}+\sqrt{\Gamma}\hat{f}_L,
\end{equation}
where $\Gamma=\alpha v_g$ is the decay time coefficient, $\alpha=\frac{2R_s}{377 h\sqrt{\frac{\mu_r}{ \epsilon_r}}}\frac{\frac{h}{W}(\frac{\omega_c}{\omega})^2+0.5}{\sqrt{1-(\frac{\omega_c}{\omega})^2}}$ denotes the attenuation coefficient, and $v_g=c\sqrt{1-(\frac{\omega_c}{\omega})^2}$ represents the group velocity. Here, $R_s=\sqrt{\frac{\omega \mu_m}{\sigma}}$, $\sigma$ and $\mu_m$ are the surface impedance, the conductivity, and the permeability of the waveguide's metal material, respectively. The quantum noise operator $\hat{f}_L(t)$ obeys to the following correlation relation: $\langle f_L(t_1)^\dagger  f_L(t_2) \rangle= N_{th} \delta(t_1-t_2)$, where $N_{th}=\frac{1}{exp(\hbar \omega /k_B T)-1}$, $\hbar$ is the plank's constant, $k_B$ represents the Boltzman constant, and $T$ denotes the waveguide temperature.

\subsection{Quantum Signal and Noise}

A quantum source hosted in a dilution refrigerator at the transmitter generates a signal with an annihilation operator $\hat{s}$. The signal is preamplified at the cryogenic temperature before being transmitted, as shown in Fig. 1. The preamplifier is coupled to the microwave waveguide with a coupling coefficient $K$. Hence, the annihilation operator of the $TE_{10}$ mode at the input port of the waveguide can be calculated using the input–output relation \cite{ours18,inputout}, given by (see Appendix A):
\begin{equation} \label{eq01}
\hat{u}=\sqrt{G}\sqrt{K}\hat{s}+\sqrt{G}\sqrt{F_a}\sqrt{K}\hat{f}_s,
\end{equation}
where $G$ and $F_a$ are the gain and the noise factor of the cryogenic preamplifier, respectively, and $\hat{f}_s$ is the source-generated noise operator. The source noise operator is characterized by zero average $\left\langle \hat{f}_s\right\rangle=0$ and associated noise photons described by $N_s=\left\langle \hat{f}_s^\dagger\hat{f}_s\right\rangle=\frac{1}{2}coth(\frac{\hbar \omega}{4 \pi k_B T_{s}})$ \cite{Clerk,noisescource}, where $T_{s}$ is the cryogenic source temperature.

By substituting the input field operator in Eq.(\ref{eq01}) into the governing equation of motion in Eq. (\ref{m5}), the field operator $\hat{u}(t)$ at the output port of the waveguide can be expressed as:  
\begin{equation} \label{eq1}
\begin{split}
&\hat{u}(\tau)=\hat{s}\sqrt{G}\sqrt{K}e^{-\frac{\Gamma}{2}\tau}+\sqrt{G}\sqrt{F_a}\sqrt{K}\hat{f}_s e^{-\frac{\Gamma}{2}\tau}\\&+\sqrt{\Gamma}e^{-\frac{\Gamma}{2}\tau} \int_{0}^\tau e^{\frac{\Gamma}{2}t}\hat{f}_L(t) dt,
\end{split}
\end{equation}
where $\tau=\frac{l}{v_g}$ is the interaction/propagation time, and $l$ is the waveguide length. Note that the noise operators $\hat{f}_s$ and $\hat{f}_L$ are uncorrelated. 

The solution in Eq. (\ref{eq1}) can be used in conjuction with the noise properties to determine the number of photons of the $TE_{10}$ mode at the output port of the waveguide (see Appendix B): 
\begin{equation} \label{eq2}
\begin{split}
\left\langle \hat{u}(\tau)^\dagger\hat{u}(\tau)\right\rangle= & \left\langle \hat{s}^\dagger\hat{s}\right\rangle G K e^{-\Gamma \tau}+G K F_a N_s e^{-\Gamma\tau}\\&+ N_{th}(1-e^{-\Gamma \tau}).
\end{split}
\end{equation}
The first term in Eq. (\ref{eq2}) is the number of $TE_{10}$ signal photons, the second term includes the source-generated noise photons and the added preamplification noise photons, and the last term is the dissipation-fluctuation induced noise photons. As can be observed from Eq. (\ref{eq2}), the noise contribution at the receiver is significant. Hence, coherent signal transmission requires implementing a scheme for noise suppression. This task is challenging without waveguide cooling to the cryogenic temperatures. 

To overcome this dilemma and suppress the noise photons yet without any waveguide cooling, we propose the following approach. A superconducting loop antenna is used inside the waveguide output port and subjected to the $TE_{10}$ flux. An \textit{LC} harmonic oscillator placed outside the waveguide is coupled to the loop antenna, as schematically demonstrated in Fig. 1. The loop antenna induces a voltage across the coupled \textit{LC} harmonic oscillator based on Faraday's law of induction. The inductance of the loop antenna, which is given by $L_a=\frac{\mu_0 \mu_{r}}{\pi} [-2(W_r+h_r )+2\sqrt{W_r^2+h_r^2 }+\varrho]$, is included within the inductance of the \textit{LC} circuit. Here, $\varrho=-h_r ln\Big(\frac{h_t+\sqrt{h_t^2+W_r^2}}{W_r}\Big)-W_r ln\Big(\frac{W_r+\sqrt{h_r^2+W_r^2}}{W_r}\Big)+h ln\Big(\frac{2 h_r}{0.5 d}\Big)+W_r ln\Big(\frac{2 W_r}{0.5 d}\Big)$, and the dimensions $d$, $W_r$ and $h_r$ are the  thickness, width, and the height of the loop antenna, respectively. The antenna's geometry and the \textit{LC} components are designed to suppress the induced photons. It then follows that the noise photons can be significantly suppressed to levels obtained in a cryogenically refrigerated waveguide. However, the number of the signal photons can be maintained intact by providing cryogenic preamplification at the transmitter. It is essential to note that the added preamplification noise photons experience the waveguide attenuation and loop antenna suppression and thus have a limited contribution. Thermal occupation of the waveguide and the dissipation-fluctuation induced noise photons are also suppressed at the received. On the contrary, while the source-generated noise photons are attenuated and suppressed, they are equally experiencing a preamplification gain. Nevertheless, the source-generated noise photons are reduced by cooling the source at the transmitter to cryogenic temperatures.  
Classically, the induces voltage across the \textit{LC} circuit is governed by the Faraday's law of induction, given by:
\begin{equation} \label{m6}
   V(t)=-\frac{\partial \Psi}{\partial t},
\end{equation}
where $\Psi= \mu_0\mu_r \int_{0}^{W_r} \int_{0}^{h_r}\vec{H}(x,y,z,t)\cdot \partial \vec{\mathbb{A}}$ is the flux to which the loop antenna is subjected and $ \vec{\partial \mathbb{A}=}\,\partial x\,\partial y \vec{e}_z$ represents the differential element of the enclosed area of the antenna. Here, $W_r$ and $h_r$ are the width and the height of the superconducting loop antenna along the $x$ and $y$ axes, respectively.

By using the expression of the field $\vec{H}$ in Eq.(\ref{m2}), the induced voltage across the \textit{LC} circuit can be described by:
\begin{equation} \label{m7}
V(t)=V_I e^{-i\omega t}+c.c.,
\end{equation}
where $V_I=i \mu_0\mu_r A h_r W_r$. The voltage in Eq. (\ref{m7}) can be quantized using the following relation:
\begin{equation} \label{m8}
   V_I=\sqrt{\frac{\hbar \omega}{C}} \hat{b},
\end{equation}
where $\hat{b}$ is the annihilation operator of the voltage in the \textit{LC} harmonic oscillator; $C$ is the capacitance, satisfying $\omega=\frac{1}{\sqrt{L C}}$; and $L$ is the inductance. The quantization relations given in Eqs. (\ref{m3}) and (\ref{m8}) can be used to obtained a direct relation between the annihilation operators of the $TE_{10}$ mode and the \textit{LC} voltage:   
\begin{equation} \label{m9}
   \hat{b}=i\hat{u} \frac{C^\frac{1}{2}}{\varphi^\frac{1}{2} (\epsilon_0\epsilon_{eff} V_{ol})^\frac{1}{2}}  \mu_0\mu_r \omega h_r W_r.
\end{equation}
Using the expressions in  Eqs. (\ref{eq1}) and (\ref{m9}), and by writing the input-output  relation at the receiver port of the waveguide (see Appendix A), the induced voltage $\hat{b}$ across the \textit{LC} circuit can be given by:
\begin{equation} \label{eq1b}
\begin{split}
&\hat{b}=\sqrt{\kappa} \sqrt{K}\sqrt{G} \hat{s} e^{-\frac{\Gamma}{2}\tau}+\sqrt{\kappa} \sqrt{K} \sqrt{G}\sqrt{F_a}\hat{f}_s e^{-\frac{\Gamma}{2}\tau}\\&+\sqrt{\kappa} \sqrt{\Gamma}e^{-\frac{\Gamma}{2}\tau} \int_{0}^\tau e^{\frac{\Gamma}{2}t}\hat{f}_L(t) dt+\sqrt{\kappa} \hat{f}_w ,
\end{split}
\end{equation}
where $ \kappa=\frac{C \omega^2 \mu_r^2 \mu_0^2 h_r^2 W_r^2}{\frac{1}{2} \Omega^2 l W h ( \epsilon_0  \epsilon_{eff} Z_F^2+ \mu_0 \mu_r)}$ is the loop antenna coupling coefficient, and $\hat{f}_{w}$ is the thermal occupation of the waveguide, which obeys the characterizations $\langle f_w \rangle= 0$ and $\langle f_w^\dagger(\tau_1)  f_w(\tau_2) \rangle= N_{th} \delta(\tau_1-\tau_2)$. It then follows that the number of induced photons across the \textit{LC} harmonic oscillator can be given by (see Appendix B): 
\begin{equation} \label{eq3}
\begin{split}
\left\langle \hat{b}^\dagger\hat{b}\right\rangle=  &\kappa K G  \left\langle \hat{s}^\dagger\hat{s}\right\rangle e^{-\Gamma \tau}+  \kappa K G F_a N_s e^{-\Gamma\tau} \\&+\kappa N_{th}(1-e^{-\Gamma \tau})+ \kappa N_{th}.
\end{split}
\end{equation}The first term in Eq. \ref{eq3} corresponds to the number of induced signal photons (denoted by $M_s$), and the sum of the second and third terms corresponds to the number of induced noise photons (denoted by $M_n$). Note that the antenna dimensions can be designed to obtain a proper value for $\kappa$ that sufficiently suppresses the induced noise photons. The signal photons can be maintained significant by providing proper gain. We note here that the main advantage of utilizing the loop antenna at the receiver, instead of a typical coax transition, is leveraging the property of geometrical control of the coupling parameter.

\section{Performance Evaluations}
A practical room temperature aluminum waveguide of conductivity $\sigma=3.5\times 10^7 S/m$ is considered in the numerical estimations presented in this work. The waveguide dimensions are designed to support single mode propagation of the fundamental $TE_{10}$ mode. For example, for a waveguide, $W=2.8$ cm width and $h=1.4$ cm height, only the $TE_{10}$ is the propagating mode for the frequency band from $6$ to $11$ GHz. The waveguide width and the $\frac{W}{h}$ ratio can be adjusted to attain desired operation bandwidth (see Appendix C).

Cryogenic amplification for the superconducting quantum circuits placed in dilution refrigerators is typically composed of two stages\cite{Engineering9}. The first stage is conducted at a cryogenic temperature of a few mK using a travelling wave parametric(TWPA) amplifier. The second stage is conducted at cryogenic temperature of a few K (e.g., 3 K) using a high-electron-mobility transistor (HEMT) amplifier. Considering TWPA and HEMT amplifiers with gains of $G_{T}$ and $G_H$, respectively, and noise factors of $F_T$ and $F_H$, respectively, the preamplification gain is $G=G_{T}. G_H$, and the noise factor is $F_a=F_{T}+\frac{F_{H}-1}{G_{T}}$. Normally, the noise factors are expressed in terms of the noise temperatures through the relation $F_{\zeta}=1+\frac{T_{\zeta}^{n}}{T_{\zeta}^{I}}$. Here, $\zeta \in \{T, H\}$ denote the TWPA and HEMT amplifiers, $T_{\zeta}^N$ denotes the noise temperature, and $T_{\zeta}^{I}$ denotes the input noise temperature. The input noise temperature of the two amplifiers is $T_{\zeta}^{I}=\frac{\hbar \omega}{4 \pi k_b} coth(\frac{\hbar \omega}{4 \pi k_B T_{\zeta}^{O}})$ \cite{noisescource}, where $T_{\zeta}^{O}$ is the corresponding physical operating temperature. The noise temperature of the TWPA amplifier is $T_T^N=\frac{\hbar \omega}{2 \pi k_b}\Big(\frac{1}{G_T ln(1+\frac{1}{G_T-1})}-\frac{1}{2}\Big)$ \cite{TWPA}, whereas the noise temperature of the HEMT amplifier is typically on the order of a few Kelvins. In this study, we consider a practical TWPA amplifier with a $G_T=10$ dB gain and an operating temperature of $T_{T}^O= 20 mK$ \cite{noisescource, TWPA}, and an off-the-shelf commercially available HEMT amplifier with a $G_H=30$ dB gain, an operating temperature of $T_H^O=10 K$, and a noise temperature of $T_H^N=4.4 k$ \cite{HEMT}. 

The signal transmittance and the output signal-to-noise ratio (across the \textit{LC} harmonic oscillator) can be obtained from Eq. (\ref{eq3}), as follows:
\begin{equation} \label{eq4}
T_r =  \kappa G e^{-\Gamma \tau}.
\end{equation}

\begin{equation} \label{eq44}
SNR_o=\frac{\left\langle \hat{s}^\dagger\hat{s}\right\rangle}{N_s F_a-\frac{N_{th}}{GK}+2\frac{N_{th} }{GK}e^{\Gamma \tau}}.
\end{equation}

\begin{figure}
\centering
\includegraphics[width=1\linewidth]{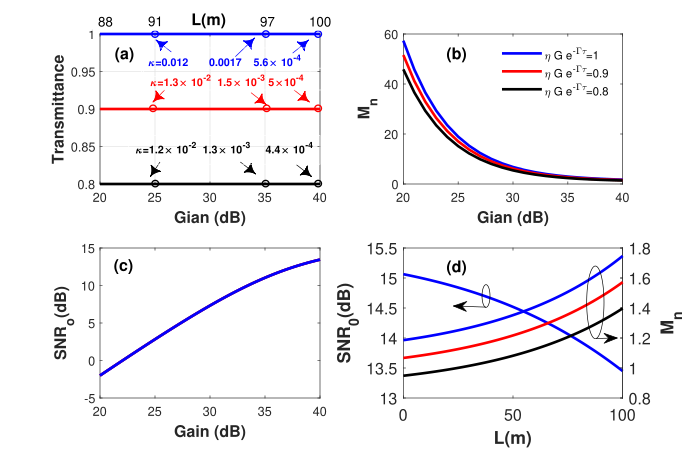}
\caption{(a) The signal transmittance versus the gain. Here, the waveguide length $l=100$ m. (b) The number of induced noise photons ($M_n$) versus the gain. Here, the waveguide length $l=100$ m. Different signal transmittance are considered. (c) The signal-to-noise ratio versus the gain. Here, the waveguide length $l=100$ m. (d) The signal-to-noise ratio and the number of induced noise photons ($M_n$) versus the waveguide length. Here, the gain $G=40$ dB. Different signal transmittances are considered.}
\label{fig:trans}
\end{figure}
Fig. 2 (a) shows the signal transmittance versus the preamplification gain for different waveguide lengths and coupling coefficient $\kappa$. Here, $\omega=10$ GHz. As can be seen, the system transmittance can be maintained unity by having proper preamplification. The number of noise photons induced across the \textit{LC} harmonic oscillator ($M_n$) is shown in Fig. 2 (b). Here, $l= 100$ m. The results show that a higher preamplification gain produces smaller induced noise photons. We note that this is provided by having a smaller $\kappa$ coefficient that maintains an intact transmittance. The output signal-to-noise ratio ($SNR_O$) is shown in Fig. 2 (c) for the same parameters and assuming $\left\langle \hat{s}^\dagger\hat{s}\right\rangle=40$ signal photons. Here, the signal-to-noise ratio is increasing with the preamplification gain, implying a mild impact of the amplification noise on the system performance. In Fig. 2 (d), $SNR_O$ and $M_n$ are plotted versus the waveguide length. Here, the same parameters are assumed for a gain of $G=40 dB$. The results show that $SNR_O$ is independent of the loop antenna parameter $\kappa$. However, a smaller $\kappa$ provides a smaller number of noise (and signal) photons. This can be explained by noting that while the thermal noise photons are intensively suppressed, by having a gain of 40 dB and $\kappa=5.6\times 10^{-4}$, the dominating contribution comes from the source-generated noise photons that are only sensitive to the transmittance. 

For completeness, we mention that the coupling parameter $\kappa=5.6\times 10^{-4}$ can be achieved with the following specifications: Capacitance of the  \textit{LC} oscillator of $C=0.5 pF$, inductance of the \textit{LC} oscillator of $L=20nH$, and loop antenna dimensions of $W_r=0.27$ mm and $h_r=0.139$ mm. The inductance of the loop antenna with these dimensions is $L_a=0.46 nH$. The inductance of the loop antenna is typically small and thus included within the inductance of \textit{LC} harmonic oscillator. 

The performance of the transmission system can be determined by calculating the noise temperature $T_{Sys}^N$, as follows:
\begin{equation} \label{eq12}
T_{Sys}^N=(F_s-1)T_{Sys}^{I},
\end{equation}
where $T_{Sys}^{I}$ represents the input noise temperature of the system (which is equal to the TWPA input noise temperature $T_T^{I}$), and $F_s=\frac{SNR_I}{SNR_O}=F_a-\frac{ N_{th}}{G K N_s}+ \frac{ 2N_{th}}{G K N_s}e^{\Gamma \tau}$ is the system noise factor, where $SNR_I=\frac{\left\langle \hat{s}^\dagger\hat{s}\right\rangle}{N_s}$ denotes the input signal-to-noise ratio at the source.

\begin{figure}
\centering
\includegraphics[width=1\linewidth]{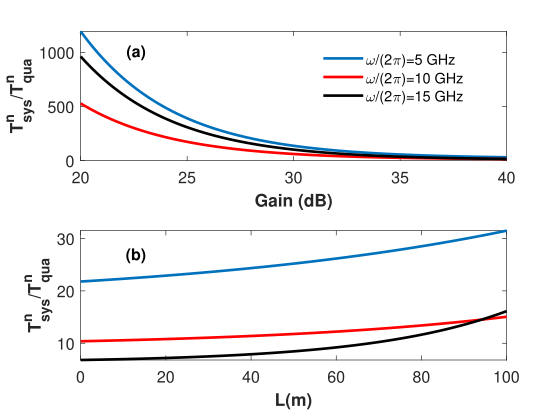}
\caption{(a) The normalized system noise temperature (with respect to the quantum-limit noise temperature) versus the gain. Here, the waveguide length $l=100$ m, the transmittance $T_r=1$, and three different signal frequencies are considered. (b) The normalized system noise temperature (with respect to the quantum-limit noise temperature) versus the waveguide length. Here, the gain $G=40$ dB, the transmittance $T_r=1$, and three different signal frequencies are considered.}
\label{fig:trans}
\end{figure}
Fig. 3 (a) shows the system noise temperature (normalized against the quantum-limit noise temperature) versus the preamplification gain. Here, the waveguide length is $l=100$ m, a unity transmittance is assumed, and $T_{Q}^N=\frac{\hbar \omega}{4 \pi k_b }$ is the quantum-limit noise temperature. Interestingly, providing proper preamplification and loop antenna design at the receiver results in near-quantum-limited transmission for the proposed system. For example, while the system noise temperature at $10$ GHz is about $10^5 (10^3)$ times the quantum-limit noise temperature for zero (20)dB preamapalification gain, the value is only $15$ times the quantum-limit noise temperature for 40 dB preamplifcation gain. This very interesting finding shows the viability of the proposed system as a quantum interconnector. The simulations in Fig.3 (b) show the normalized system noise temperature versus the waveguide length while considering the same parameters as in Fig. 3 (a) and a premplification gain of 40 dB. One can see that the system noise temperature is higher for lower frequencies. This is an expected observation as the thermal noise occupation is inversely proportional with frequency. However, counterintuitively, the system noise temperature at $15$ GHz exceeds the system's noise temperature at $10$ GHz for waveguide lengths more than 80 m. We note here that the attenuation factor for $15$ GHz is greater than that of the $10$ GHz. Hence, for a large enough waveguide length, the propagation losses are significant, causing the noise temperature of the  $15$ GHz to surpass that of the $10$ GHz noise temperate.   

To further evaluate the system performance, we consider a single quantum bit of information that is generated by the cryogenic source in Fig. 1 and launched toward the room-temperature waveguide for transmission through the proposed system. The generated quantum state at the source is given by\cite{state1,state2}:
\begin{equation} \label{eq5}
|\psi_s\rangle=a |0\rangle+b \int_{-\infty}^{+\infty}d \omega \xi^*(\omega)|1_\xi\rangle,
\end{equation}
where $|0\rangle$ and $|1_\xi\rangle=\hat{s}^\dagger|0\rangle$ are the vacuum  and wave packet states, respectively, $|a|^2+|b|^2=1$, and $\xi(\omega)$ is the spectral profile of the generated wave packet.
The source-generated thermal noise $\hat{f}_s$ is uncorrelated with the signal $\hat{s}$. Hence, the corresponding source density matrix is given by:
\begin{equation} \label{eq6}
\begin{split}
\rho_s=&|a|^2 |0\rangle \langle0|+|b|^2 \int_{-\infty}^{+\infty}d\omega^{'} \int_{-\infty}^{+\infty}d\omega \xi^*(\omega^{'})\xi(\omega) |1_\xi\rangle \langle1_\xi|\\&+ab^*\int_{-\infty}^{+\infty}d \omega^{'} \xi(\omega^{'})|0\rangle \langle1_\xi|+ba^*\int_{-\infty}^{+\infty}d \omega \xi^*(\omega)|1_\xi\rangle \langle0|\\&+\int_{-\infty}^{+\infty}d\omega^{'} \int_{-\infty}^{+\infty}d\omega \hat{f}_s^\dagger(\omega^{'})\hat{f}_s(\omega) |0\rangle \langle0|.
\end{split}
\end{equation} 
The output density operator of the corresponding quantum state across the \textit{LC} harmonic oscillator is obtained by following the same steps implemented in the previous session to derive Eqs.(\ref{eq1b}) and (\ref{eq3}), yielding: 
\begin{equation} \label{eq8}
\begin{split}
&\rho_{LC}=|a|^2\kappa K G e^{-\Gamma \tau} |0\rangle\langle0|+|b|^2 \kappa K G e^{-\Gamma \tau} |1_\xi\rangle \langle1_\xi|\\&+ ab^* \kappa K G e^{-\Gamma \tau} \int_{-\infty}^{+\infty}d \omega \xi(\omega) |0\rangle\langle1_\xi|\\&+ba^* \kappa K G e^{-\Gamma \tau} \int_{-\infty}^{+\infty}d \omega \xi^*(\omega) |1_\xi\rangle\langle0|\\&+ \kappa K  G F_a e^{-\Gamma \tau} \int_{-\infty}^{+\infty}d\omega^{'} \int_{-\infty}^{+\infty}d\omega \hat{F}_s^\dagger(\omega^{'})\hat{F}_s(\omega) |0\rangle\langle0|\\& +\kappa (1-e^{-\Gamma \tau}) \int_{-\infty}^{+\infty}d\omega^{'} \int_{-\infty}^{+\infty}d\omega \hat{F}_L^\dagger(\omega^{'})\hat{F}_L(\omega)|0\rangle \langle0|\\&+\kappa \int_{-\infty}^{+\infty}d\omega^{'} \int_{-\infty}^{+\infty}d\omega \hat{F}_w^\dagger(\omega^{'})\hat{F}_w(\omega) |0\rangle \langle0|,
\end{split}
\end{equation}
where $\hat{F}_{\iota}(\omega)$ is the frequency spectrum of the $\hat{f}_{\iota}(t)$ operator, obeying  $\langle F_\iota^\dagger(\omega^{'})  F_\iota(\omega) \rangle= 2\pi N_{\iota} \delta(\omega^{'}+\omega)$, and $\iota \in{\{s,L,w\}}$. The first two terms in Eq. (\ref{eq8}) correspond to the qubit information generated by the source, the third term corresponds to the source-generated noise and the preamplification noise, the fourth term corresponds to the fluctuation-dissipation generated noise, and the last term corresponds to the thermal occupation noise of the waveguide. 

The closeness of the input and output quantum states in our proposed transmission systems can be measured by calculating the fidelity between the quantum state at the source and the quantum state across the \textit{LC} harmonic oscillator at the receiver (which is hereafter refereed to as the transmission fidelity $D_t$) as follows \cite{fidelity}:
\begin{equation} \label{eq9}
D_t=\frac{tr(\rho_s\rho_{LC})}{\sqrt{tr(\rho_s^2)}\sqrt{tr(\rho_{LC}^2)}},
\end{equation}
where $tr()$ is the trace operator.

Fig. 4 presents the transmission fidelity (averaged over all possible Bloch sphere states). Here, a transmittance of unity and three different microwave frequencies are considered. 
\begin{figure}
\centering
\includegraphics[width=1\linewidth]{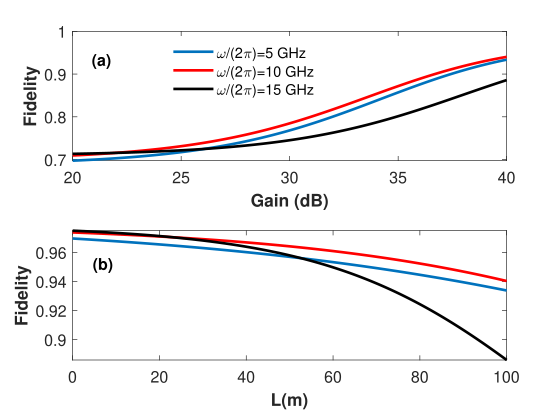}
\caption{(a) The transmission fidelity $D_t$ versus the gain. A waveguide length $l=100$ m and a transmittance of unity are considered. (b) The transmission fidelity $D_t$ versus the waveguide length. A gain $G=40$ and a transmittance of unity are considered.}
\label{fig:frogg}
\end{figure}
In Fig. 4 (a) the fidelity is presented versus the preamplification gain. Interestingly, suitable preamplification can produce fidelity close to unity (greater than $95\%$) for waveguide lengths up to one hundred meters. As expected, it can be observed that the fidelity is increasing for higher frequencies. However, for gain ranges greater than 25 dB, the fidelity for the $5$ and $10$ GHz frequencies are showing greater values than that of the  $15$ GHz. We note that for low gain ranges, the suppression at the receiver is partial ( i.e., a small gain designates a large $\kappa$ ), leading to thermally limited operation. In contrast,  large gain ranges lead to substantial suppression (owing to the associated small $\kappa$ values) causing a suppression limited operation. Nevertheless, higher frequencies are suffering from greater attenuation. Thus, for a unity transmittance, a higher frequency requires relatively smaller $\kappa$. Fig. 4(b), presents the fidelity versus the waveguide length. A preamplification gain of 40 dB is considered. One can observe that fidelity is thermally-limited for low waveguide lengths less than 29 m, and suppression-limited for waveguide lengths greater than 60 m. Here, same notes can be mentioned regarding the thermal occupation and the suppression dominance for different frequencies. The simulations in Fig. 4 suggest that the preamplification noise is of a mild impact (i.e., increasing gain results in a higher fidelity). We explain this by noting that the preamplification noise photons are prone to waveguide attenuation and receiver suppression. Finally, it is worthy to point out that the source-generated noise photons are proportional to the transmittance, and thus their impact is mitigated only by cooling the transmitter.
\section{Discussion}
In quantum microwave transmission systems, the transmission fidelity is severely degraded at room temperature due to significant thermal occupation within the microwave frequency spectrum. Thus, a waveguide cooled to cryogenic temperatures has been proposed to mitigate thermal occupation and achieve acceptable transmission quality. Other techniques, such as implementing controllable coupling between qubits (at the two sides of the link) and the connecting channel, have also been proposed. The potentials and limitations of these techniques were briefly discussed in the introduction.

In this study, we propose a novel technique to achieve reduced noise levels without waveguides refrigeration simply by designing a suitable loop antenna at the receiver side. For example, in Fig 4, the thermal waveguide occupation at room temperature is $N_{th}= 1.22\times 10^3$ for a microwave frequency of $\frac{\omega}{2 \pi}=5$ GHz. However, the number of the induced noise photons across the \textit{LC} harmonic oscillator is only $\kappa N_{th}= 0.413 $, which is equivalent to that for the same waveguide cooled to $0.2$ K. Same calculations for $\omega=10$ GHz, with $N_{th}= 609$ and $\kappa N_{th}= 0.34 $, corresponds to noise level equivalent to a cooled waveguide at $0.35$ K. Combining this proposed loop antenna technique with proper cryogenic preamplification results in a room-temperature lossy waveguide with high-fidelity transmission (above $95\%$) over significant transmission distances (up to $100$ m). 

To the best of our knowledge, this study is the first proposal of a high-fidelity microwave transmission system using a typical lossy nonrefrigerated waveguide. Our approach has the potential to realize a modular quantum computer with waveguides placed outside dilution refrigerators. This feature is very important because scaling quantum computers up to a capacity of thousands of qubits is crucial for leveraging quantum supremacy.  The state-of-the-art capacity of current superconducting quantum computers is less than $127$ qubits \cite{supremacy17,IBM23,processor27,IBM127}, and boosting the number of qubits to thousands is very challenging, especially from a heat management standpoint. This difficulty is due to the fact that adding a qubit requires connecting a considerable number of cables and related components, which imposes an overwhelming heating load. For example, the estimated cost per qubit to maintain the required cryogenic refrigeration and connect the pertinent cables is approximately $\$10$ K\cite{race28}. Thus, a scaled quantum computer utilizing a giant dilution refrigerator is expected to cost tens of billions of dollars \cite{million29}. Connecting separated quantum nodes (or processors) by coherent signaling is a promising approach for efficient scaled quantum computation. The findings reported here have the potential to expedite the realization of sought-after modular superconducting quantum computers capable of containing thousands (million) of qubits.
Finally, we note that the proposed scheme (cryogenic preamplification at the transmitter and a passive cryogenic detection circuit at the receiver) is technically very easy to implement.

\section{Methods}
\subsection{The input-output relations}
\begin{figure}[tbhp]
\centering
\includegraphics[width=1\linewidth]{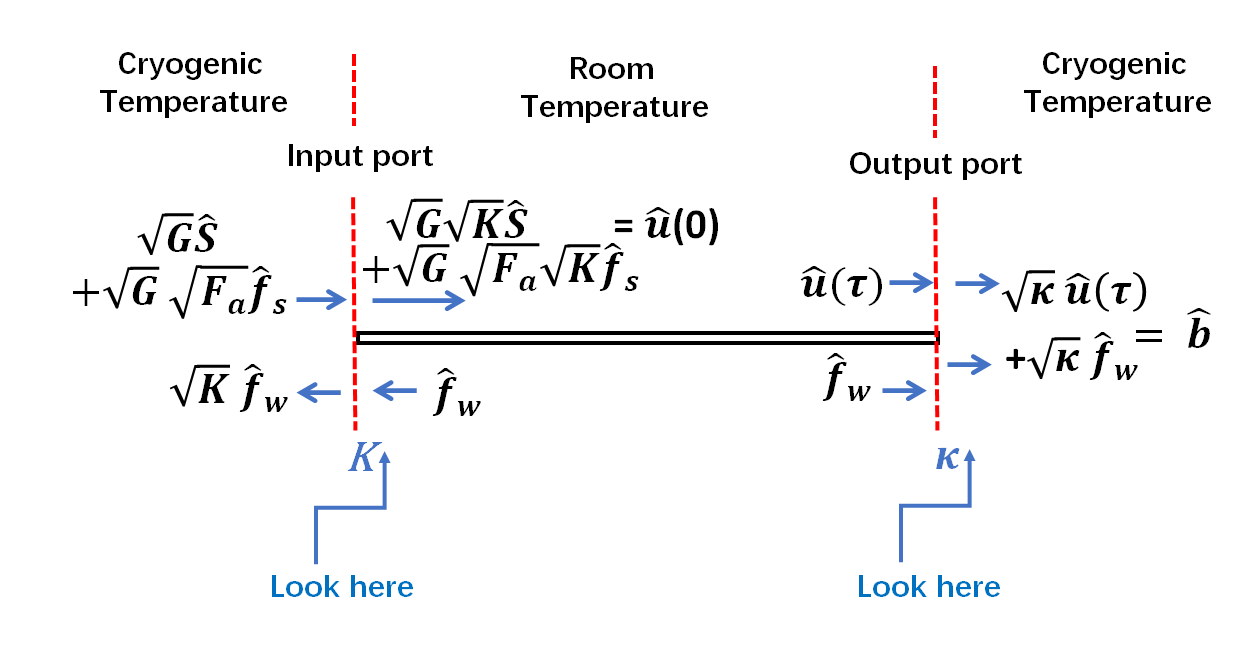}
\caption{The fields operators and the input-output relations at the waveguide ports. }
\label{App1}
\end{figure}

The fields operators at the waveguide ports can be described through the input-out relations. The schematic in Fig. \ref{App1} shows the fields operators at the two ports of the waveguide with the corresponding coupling parameters $K$ and $\kappa$. It can be seen that the thermal occupation of the waveguide loads the transmitter and receiver nodes with $\sqrt{K} \hat{f}_w$ and $\sqrt{\kappa} \hat{f}_w$ operators, respectively. The thermal load at the transmitter can be extracted by implementing a circulator between the amplifier and the waveguide. However, the thermal load at the receiver can be suppressed by controlling the coupling parameter $\kappa$. The operators' input-output relations impose:  $\hat{u}=\sqrt{G}\sqrt{K}\hat{s}+\sqrt{G}\sqrt{F}\sqrt{K}\hat{f}_s$ at the input of the waveguide, and  $\hat{b}= \sqrt{\kappa} \hat{u}(\tau) +\sqrt{\kappa} \hat{f}_w $ at the output port of the loop antenna, where  $\hat{u}(\tau)$ is given by Eq.(\ref{eq1}).

\begin{figure}[tbhp]
\centering
\includegraphics[width=1\linewidth]{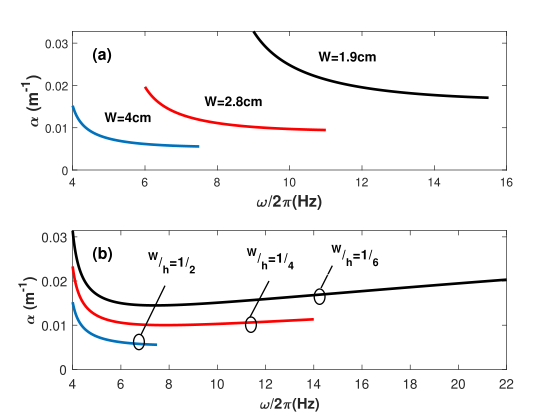}
\caption{The attenuation factor versus the single-mode-bandwidth for  $TE_{10}$ mode. (a) The attenuation factor versus the single-mode bandwidth for $W/h=1/2$. Different waveguide widths are considered. (b) The attenuation factor versus the single-mode bandwidth for $W=2.8 cm$. Different $W/h$ ratios are considered. }
\label{App2}
\end{figure}

\subsection{Number of photons}
Using the field operator $\hat{u}(t)$ expression in Eq. (\ref{eq1}), and noting that the noise operators $\hat{f}_s$ and $\hat{f}_L(t)$ are uncorrelated, the number of photons at the output of the rectangular waveguide is given by: 
\begin{equation} 
\begin{split}\label{num}
&\left\langle \hat{u}(\tau)^\dagger\hat{u}(\tau)\right\rangle=  \left\langle \hat{s}^\dagger\hat{s}\right\rangle K G e^{-\Gamma \tau}+K G F_a \left\langle \hat{f}_s^\dagger\hat{f}_s\right\rangle e^{-\Gamma\tau}\\&+\Gamma e^{-\Gamma\tau} \int_{0}^\tau e^{\frac{\Gamma}{2}t^{'}} \int_{0}^\tau e^{\frac{\Gamma}{2}t}\left\langle \hat{f}^\dagger_L(t^{'})\hat{f}_L(t)\right\rangle dt dt^{'} .
\end{split}
\end{equation}
The expression of the number of photons in Eq.(\ref{num}) can be simplified by incorporating the noise properties, which are $\langle f_s^\dagger(t^{'}) f_s(t) \rangle= N_s \delta(t^{'}-t)$ and $\langle f_L^\dagger(t^{''}) f_L(t) \rangle= N_{th} \delta(t^{''}-t)$, yielding the form in Eq.\ref{eq2}.

\subsection{Waveguide dimensions}
The waveguide dimensions are designed to support a single mode operation. The cut-off frequency of the propagating modes in a rectangular microwave waveguide is given by $\omega^{mn}_c=\frac{2 \pi c}{\sqrt{\epsilon_r \mu_r}} \sqrt{(\frac{m}{W})^2+(\frac{n}{h})^2}$, where $n=1,2,3,...$ and $m=1,2,3,...$. The lowest cut-off frequency is associated with the $TE_{10}$ mode, given by $\omega^{10}_c=\frac{2 \pi c}{2W\sqrt{\epsilon_r}}$. The second lowest cut-off frequency is for the $TE_{01}$ mode, given by $\omega^{01}_c=\frac{2 \pi c}{2h\sqrt{\epsilon_r}}$. Hence, the single mode frequency range (named hereafter single mode bandwidth) is given by $B_w=\omega^{10}_c-\omega^{01}_c=\frac{2 \pi c}{2\sqrt{\epsilon_r \mu_r}}\frac{W-h}{W h}$.

Fig.6 shows the attenuation factor along with the single mode bandwidth for $TE_{10}$ mode. In Fig. \ref{App2} (a), the simulations are shown considering different waveguide widths for $\frac{W}{h}=\frac{1}{2}$. On the other hand, Fig. \ref{App2}(b) considers $W=2.8$ cm for different $\frac{W}{h}$ ratios. It can be seen that smaller waveguide widths result in larger single-mode bandwidths. However, smaller widths endure larger mode attenuation. Furthermore, for a given waveguide width, the smaller the  $\frac{W}{h}$ ratio, the larger the single-mode bandwidth. Yet, the experienced attenuation is increasing. Hence, a trade-off takes place between the single-mode bandwidth and the waveguide attenuation.

\section*{Data Availability}
The datasets generated during and/or analysed during the current study are available from the corresponding author on reasonable request.

\section*{Acknowledgements}
This research is supported by Abu Dhabi Award for Research Excellence under ASPIRE/Advanced Technology Research Council (AARE19-062) 2019.
\section*{Author Contributions}
M.Q. conceived the idea, developed the numerical model, prepared the presented results, and took the lead in writing the manuscript. H.E. contributed to the developed model, validated the obtained results, and contributed to the manuscript writing.

\section*{Competing Interests} The Authors declare no Competing Financial or Non-Financial Interests.
\section*{Correspondence} 
Correspondence and requests for materials
should be addressed to Montsir Qasymeh~(email: montasir.qasymeh@adu.ac.ae).

\begin{thebibliography}{99}
%
\bibitem{logic1} S. Daiss et al., "A quantum-logic gate between distant quantum-network modules,"Science{\bf 371}, no. 6529, pp. 614 (2021). 

\bibitem{Deterministic2} Y. Zhong, "Deterministic multi-qubit entanglement in a quantum network,"Nature{\bf 571} (2021). 

\bibitem{Microwave3} P. Magnard et al., "Microwave Quantum Link between Superconducting Circuits Housed in Spatially Separated Cryogenic Systems,"Phys. Rev. Lett.{\bf 125}, no. 260502 (2020). 


\bibitem{coherent4} J.G. Bartholomew et al., "On-chip coherent microwave-to-optical transduction mediated by ytterbium in YVO4,"Nat. Commun{\bf 11}, no. 3266 (2020). 

 
\bibitem{Demonstrating5} D. Gottesman and I. Chuang, "Demonstrating the viability of universal quantum computation using teleportation and single-qubit operations,"Nature{\bf 402}, pp.390--393 (1999). 

\bibitem{information6} T. Northup and R. Blatt, "Quantum information transfer using photons,"Nature Photon{\bf 8}, pp.356--363 (2014). 


\bibitem{Distributed7} L. Jiang and J. Taylor and A. Sørensen and M. Lukin, "Distributed quantum computation based on small quantum registers,"Phys. Rev. A{\bf 76}, no. 062323 (2007). 

\bibitem{modular8} C. Monroe et al., "Large-scale modular quantum-computer architecture with atomic memory and photonic interconnects,"Phys. Rev. A{\bf 89}, no.022317 (2014). 


\bibitem{Engineering9} S. Krinner et al., "Engineering cryogenic setups for 100-qubit scale superconducting circuit systems,"EPJ Quantum Technol{\bf 6}, no. 2 (2019).

\bibitem{computers10} T. D. Ladd and F. Jelezko and R. Laflamme and Y. Nakamura and C. Monroe and J. OBrien, "Quantum computers,"Nature{\bf 464}, pp.45--53 (2010). 

\bibitem{Interconnects11} D. Awschalom et al., "Development of quantum interconnects (QuICs) for next-generation information technologies,"PRX Quantum{\bf 2}, no. 017002 (2021).

\bibitem{internet12} H. J. Kimble, "The quantum internet,"Nature{\bf 453}, pp.1023--1030 (2008). 


\bibitem{internet13} S. Wehner and D. Elkouss and Ronald Hanson, "Quantum internet: A vision for the road ahead,"Science{\bf 362}, no.6412 (2018). 
 
\bibitem{Entanglement14} L. Stephenson et al., "High-Rate, High-Fidelity Entanglement of Qubits Across an Elementary Quantum Network,"Phys. Rev. Lett.{\bf 124}, no.110501 (2019). 

\bibitem{Sensing15} Z. Zhang and S. Mouradian and F. Wong and J. Shapiro, "Entanglement-Enhanced Sensing in a Lossy and Noisy Environment,"Phys.Rev. Lett.{\bf 114}, no. 110506 (2015).


\bibitem{Josephson16} J. Mooij and T. Orlando and L. Levitov and L. Tian and C. van der Wal and S. Lloyd, "Josephson Persistent-Current Qubit,"Science{\bf 285}, no.5430, pp.1036--1039 (1999). 

\bibitem{supremacy17} F. Arute and K. Arya and  R. Babbush and et al., "Quantum supremacy using a programmable superconducting processor,"Nature{\bf 574}, pp.505--510 (2019). 

\bibitem{ours18} M. Qasymeh and H. Eleuch, "Hybrid two-mode squeezing of microwave and optical fields using optically pumped graphene layers,"Sci. Rep.{\bf 10}, no. 16676 (2020).

\bibitem{memories19} C. Axline et al., "On-demand quantum state transfer and entanglement between remote microwave cavity memories,"Nature Phys{\bf 14}, pp.705--710 (2018). 


\bibitem{Deterministic20} P. Kurpiers et al., "Deterministic quantum state transfer and remote entanglement using microwave photons,"Nature {\bf 558}, pp.264--267 (2018). 



\bibitem{Deterministic21} Y. Zhong et al., "Deterministic multi-qubit entanglement in a quantum network,"Nature {\bf 590}, pp.571--575 (2021). 


\bibitem{quantum22} A. Bienfait et al., "Phonon-mediated quantum state transfer and remote qubit entanglement,"Science {\bf 364}, ni. 6438, pp.368--371 (2019). 


\bibitem{IBM23} J. Gambetta, "IBM’s Roadmap For Scaling Quantum Technology,"IBM Research Blog (2020). 


\bibitem{thermalch1} B. Vermersch and P.-O. Guimond and P. Zoller, "Quantum state transfer via noisy Photonic and phononic waveguides,"Phys. Rev. Lett {\bf 118}, no.133601 (2017). 

\bibitem{thermalch2} Z.-L. Xiang and M. Zhang and L. Jiang and P. Rabl, "Intracity quantum communication via thermal microwave networks,"Phys.Rev. X.{\bf 7}, no. 011035 (2017).

\bibitem{our24}M. Qasymeh and H. Eleuch, "Quantum microwave-to-optical conversion in electrically driven multilayer graphene,"Opt. Express {\bf 27}, no.5, pp.5945-5960 (2019).

\bibitem{patent25} M. Qasymeh and H. El Euch, "Frequency-Tunable quantum microwave to optical conversion system,"U.S. Patent, no.10,824,048 B2 (2020). 

\bibitem{Superconducting26} M. Mirhosseini and A. Sipahigil and M. Kalaee and O. Painter, "Superconducting qubit to optical photon transduction,"Nature {\bf 588}, pp.599--603 (2020).

\bibitem{Pozar32} D. M. Pozar, "Microwave Engineering,"John Wiley and Sons,New York, (2012).

\bibitem{inputout} C. W. Gardiner and M. J. Collett, "Input and output in damped quantum systems: Quantum stochastic differential equations and the master equation,"Phys. Rev. A {\bf 31}, no.6, pp.3761--3774 (1980).


\bibitem{Clerk} A. Clerknet M. H. Devoret et al., "Introduction to quantum noise, measurement, and amplification,"Rev. Mod. Phys. {\bf 82}, pp.1155–1208 (2010).


\bibitem{noisescource} S. Simbierowicz et al. "Characterizing cryogenic amplifiers with a matched temperature-variable,"Review of Scientific Instruments {\bf 92}, no.3, pp.034708 (2021).

\bibitem{TWPA} L. Fasolo et al., "Bimodal Approach for Noise Figures of Merit Evaluation in Quantum-Limited Josephson Traveling Wave Parametric Amplifiers,"arXiv:2109.14924v1 (2021).

\bibitem{HEMT} F. Heinz and F. Thome and A. Leuther and O. Ambacher, "A 50-nm Gate-Length Metamorphic HEMT Technology Optimized for Cryogenic Ultra-Low-Noise Operation,"IEEE Transactions on Microwave Theory and Techn. {\bf 69}, no. 8 pp.3896--3907 (2021).

\bibitem{state1} C. Lee M. Tame J. Lim and J. Lee, "Quantum plasmonics with a metal nanoparticle array,"Phys. Rev. A {\bf 85}, no.063823 (2013).

\bibitem{state2} Z. Y. Ou, "Temporal distinguishability of an N-photon state and its characterization by quantum interference,"Phys. Rev. A {\bf 74}, no.063808 (2006).

\bibitem{fidelity} X. Wang and C.-S. Yu and X.X. Yi, "An alternative quantum fidelity for mized states of qudits,"Phys. Lett. A {\bf 373}, pp.58--60 (2008).

\bibitem{processor27}M. Gong et al., "Quantum walks on a programmable two-dimensional 62-qubit superconducting processor,"Science,  {\bf xx}, no. x (2021).

\bibitem{IBM127} B. Yirka, "IBM announces development of 127-qubit quantum processor,"Phys.org,  Nov 16 (2021).

\bibitem{race28} G. Lichfield, "Inside the race to build the best quantum computer on Earth,"MIT Technology Review, (2021).


\bibitem{million29} J. Levy, "1 million qubit quantum computers: moving beyond the current “brute force” strategy,"SEEQC, (2020).


\bibitem{Josephson30} C. Macklin et al., "A near quantum-limited Josephson traveling-wave parametric amplifier,"Science {\bf 350}, n0. 6258, pp.307--310 (2015).

\bibitem{Parametric31} U. Mendes et al., "Parametric amplification and squeezing with an ac- and dc-voltage biased superconducting junction,"Phys. Rev. Applied {\bf 11}, no. 3,  pp.034035 (2019).
\end{thebibliography}
\end{document}